\documentclass[showpacs,amsmath,amssymb,twocolumn]{revtex4}
\usepackage{graphicx}
\usepackage{dcolumn}
\usepackage{bm}

\begin{document}

\title{\bf\large{Finite Temperature Phase Diagram of a Two--Component Fermi Gas with Density Imbalance}}
\author{\normalsize{Lianyi He\footnote{Email address: hely04@mails.tsinghua.edu.cn}, Meng Jin\footnote{Email address: jin-m@mail.tsinghua.edu.cn
        } and Pengfei Zhuang\footnote{Email address:
        zhuangpf@mail.tsinghua.edu.cn}}}
\affiliation{Physics Department, Tsinghua University, Beijing
100084, China}

\begin{abstract}
We investigated possible superfluid phases at finite temperature
in a two-component Fermi gas with density imbalance. In the frame
of a general four-fermion interaction theory, we solved in the BCS
region the gap equations for the pairing gap and pairing momentum
under the restriction of fixed number densities, and analyzed the
stability of different phases by calculating the superfluid
density and number susceptibilities. The homogeneous superfluid is
stable only at high temperature and low number asymmetry, the
inhomogeneous LOFF survives at low temperature and high number
asymmetry, and in between them there exists another possible
inhomogeneous phase, that of phase separation. The critical
temperatures and the orders of the phase transitions among the
superfluid phases and normal phase are calculated analytically and
numerically. The phase diagram we obtained in the temperature and
number asymmetry plane is quite different from the one in
temperature and chemical potential difference plane for a system
with fixed chemical potentials.
\end{abstract}


\maketitle

\section {Introduction}
\label{s1}
Recently, the superfluidity in a two-component Fermi gas with
population imbalance promoted great interest in both experimental
and theoretical studies. The key point is the ground state of
fermion pairing between different species with mismatched Fermi
surfaces, and it is directly related to the studies of atomic
Fermi gas with population
imbalance\cite{caldas,petit,pao,son,sheehy,yang,dukelsky,torma,silva,aurel,hu,zhai,gu,duan,machida,massimo,pieri,mizhushima},
superconductivity with Zeeman
splitting\cite{sarma,larkin,fulde,takada}, isospin asymmetric
nuclear matter\cite{sedrakian,Isayev,sedrakian2} and dense QCD
matter\cite{huang,shovkovy,alford,abuki,pion}. In the
investigation, two exotic states, the homogeneous and isotropic
Sarma\cite{sarma} or breached pairing\cite{liu,forbes} state and
the inhomogeneous and anisotropic Larkin-Ovchinnikov-Fudde-Ferrell
(LOFF) \cite{larkin,fulde,takada,casalbuoni} state, are especially
concerned. While these two interesting states have never been
observed experimentally, the recent progress in the study of
atomic Fermi gas\cite{exp1,exp2,exp3} may provide us a way to
realize and observe them.

There are two crucial problems related to the Sarma state due to
the existence of gapless fermionic excitation. One problem is its
thermodynamical instability compared with the fully gapped BCS
state, when the effective chemical potentials for the two species
are fixed. This is called Sarma instability\cite{sarma}. It is now
accepted that the Sarma instability can be avoided in systems
under some physical constraints such as fixed particle
numbers\cite{sedrakian,liao} instead of fixed chemical potentials,
long range gauge interaction with charge neutrality
requirement\cite{shovkovy}, a proper finite range attractive
interaction\cite{forbes}, and in the strong coupling BEC
region\cite{pao,smit}. The second problem for the Sarma state is
its negative superfluid density\cite{pao,wu} and negative number
susceptibility\cite{pao}. The former leads to negative Meissner
mass squared\cite{huang3,he} for charged systems and indicates
that the LOFF state is probably energetically favored than the
Sarma state\cite{giannakis,he2}, and the latter shows that the
superfluid-normal phase separation\cite{caldas} is also more
stable than the Sarma state.

The above discussed negative superfluid density and number
susceptibility are obtained in weak coupling BCS region and at
zero temperature. Recently, it is argued that the Sarma phase will
be free from negative superfluid density and number susceptibility
in strong coupling BEC region\cite{pao,sheehy,smit}, and both
experimental and theoretical studies support that the
normal-superfluid phase separation is energetically favored around
the unitary region\cite{sheehy,exp3}.

In this paper we are interested in the BCS region where the LOFF
phase is shown to be energetically favored\cite{sheehy,hu}. An
important and interesting phenomenon in this region is the
enhanced pairing correlation at finite
temperature\cite{chen,sedrakian3}. The pairing gap is not a
monotonous function of temperature and its maximum is located not
at zero but at finite temperature. Especially, for a large number
asymmetry, the superfluidity appears only at finite temperature.
This strange and interesting phenomenon was found in the studies
of isospin asymmetric nuclear matter\cite{sedrakian,Isayev}, two
flavor color-superconducting quark matter\cite{shovkovy}, and
general breached pairing superfluidity\cite{liao}. However, it is
recently found that the superfluid density of the homogeneous
Sarma phase is negative at low temperature and becomes positive
only in a temperature window near the critical temperature
$T_c$\cite{chen,mart}. As a result, inhomogeneous phases such as
LOFF phase can enter the low temperature region, and the phase
structure should be re investigated.

The purpose of this paper is to analyze the phase structure of a
two-component Fermi gas with density imbalance in the BCS regime.
Our analysis is based on the mean field approximation which is
believed to be a good treatment in the BCS regime even at finite
temperature. We take LOFF phase into account. For the sake of
simplicity, we consider the simplest pattern of LOFF phase,
namely, the single plane wave LOFF state or the so-called FF
state. While the consideration of a general LOFF state is
complicated, we will argue that the phase structure with the FF
state will not be changed qualitatively. Due to the thermal
excitation, we will not distinguish between the gapless and gapped
phases and simply call them homogeneous superfluid (HS).

After a brief review of the theoretical frame for the attractive
two-component Fermi gas with population imbalance in Section
\ref{s2}, we calculate in Section \ref{s3} the superfluid density
and number susceptibility as a function of temperature at fixed
population imbalance and obtain the turning temperature where the
superfluid density and number susceptibility both change sign. In
Section \ref{s4} we include the LOFF state and separate the
homogeneous region from the possible LOFF region. By calculating
in Section \ref{s5} the number susceptibility $\chi$ for the LOFF
state, we further distinguish the stable LOFF region at high
population imbalance from unstable LOFF region at low population
imbalance. We obtain the phase diagram in the temperature and
population imbalance plane in Section \ref{s5}. We summarize in
Section \ref{s6}. The natural unit of $c=\hbar=k_B=1$ is adopted
through the paper.

\section {Formalism}
\label{s2}
The physical system we are interested in in this paper is an
infinite system composed of two species of fermions with
attractive interaction in three dimensional free space. Generally,
the system can be modelled by the Lagrangian density
\begin{equation}
{\cal
L}=\sum_{\sigma=\uparrow,\downarrow}\psi_{\sigma}^\dagger\left(i\partial_
t+\frac{\nabla^2}{2m}+\mu_\sigma\right)\psi_{\sigma}+g\psi_{\uparrow}^\dagger\psi_{\downarrow}^\dagger\psi_{\downarrow}\psi_{\uparrow},
\end{equation}
where $\psi_\sigma(x)$ are fermion fields for the two species
denoted by $\uparrow$ and $\downarrow$ with space-time $x=(t,{\bf
x})$, $g$ is the coupling constant, $m$ is the fermion mass, and
$\mu_\uparrow$ and $\mu_\downarrow$ are the chemical potentials.

For attractive coupling $g$ we can perform an exact
Stratonovich-Hubbard transformation to introduce the pair field
$\Phi\sim g\psi_\downarrow\psi_\uparrow$ and its complex conjugate
$\Phi^*\sim g\psi^\dagger_\uparrow \psi^\dagger_\downarrow$. With
the Nambu-Gorkov field defined as $\Psi=(\psi_\uparrow,
\psi^\dagger_\downarrow)^T$, the partition function can be
expressed as
\begin{equation}
Z=\int[d\Psi^\dagger][d\Psi][d\Phi^*][d\Phi]e^{\int_0^{\beta}
d\tau\int d^3{\bf x} \left(\Psi^\dagger{\cal
K}\Psi-\Phi^*\Phi/g\right)}
\end{equation}
in the imaginary time ($\tau=it$) formalism of finite temperature
field theory, where $\beta$ is the inverse temperature,
$\beta=1/T$, and the kernel ${\cal K}$ is defined as
\begin{equation}
{\cal K}[\Phi^*,\Phi]=\left(\begin{array}{cc}
-\partial_\tau+\frac{\nabla^2}{2m}+\mu_\uparrow&\Phi(\tau,{\bf x})
\\ \Phi^*(\tau,{\bf x})&-\partial_\tau-\frac{\nabla^2}{2m}-\mu_\downarrow\end{array}\right)\ .
\end{equation}
Integrating out the fermionic degrees of freedom, we obtain
\begin{equation}
Z=\int[d\Phi^*][d\Phi]e^{-S_{eff}[\Phi^*,\Phi]}
\end{equation}
with the effective action
\begin{equation}
S_{eff}[\Phi^*,\Phi]=\int_0^{\beta} d\tau\int d^3{\bf
x}\frac{\Phi^*\Phi}{g}-\text{Tr}\ln{\cal K}[\Phi^*,\Phi].
\end{equation}

For a dilute gas, we can replace the bare coupling constant $g$ by
the low energy limit of the two-body T-matrix\cite{pao},
\begin{equation}
\frac{m}{4\pi a_s}=-\frac{1}{g}+\int\frac{d^3{\bf
p}}{(2\pi)^3}\frac{1}{2\epsilon_p}
\end{equation}
with the $s$-wave scattering length $a_s$ and fermion energy
$\epsilon_p={\bf p}^2/(2m)$.

For an arbitrary scattering length $a_s$ at finite temperature
$T$, we should take into account the contribution from the pair
fluctuations and pseudogap\cite{chen} to the thermodynamics of the
system. Since in this paper we focus on the BCS region where the
scattering length is negative and small and the coupling is
relatively weak, the effect of pair fluctuations and pseudogap can
be approximately neglected, and the mean field approximation is a
good approach to investigate the phase structure. In the mean
field approximation, we replace the pair field $\Phi$ and its
complex conjugate by their expectation values. To have a unified
treatment for both the homogeneous and LOFF superfluid, the order
parameter for the superfluid can be defined as
\begin{equation}
\label{q}
\langle\Phi(x)\rangle=\Delta e^{2i{\bf q}\cdot{\bf x}}\
,\ \ \ \langle\Phi^*(x)\rangle=\Delta e^{-2i{\bf q}\cdot{\bf x}},
\end{equation}
where $\Delta$ is the amplitude of the order parameter and can be
taken to be real, and $2{\bf q}$ is the pair momentum in single
plane wave LOFF state. Obviously, ${\bf q}=0$ and ${\bf q}\neq 0$
correspond, respectively, to the HS and LOFF states.

Since a general LOFF state can be considered as a superposition of
single plane wave LOFF states, it corresponds to a deeper minimum
of the free energy of the system, compared with the single plane
wave LOFF state. Therefore, a general LOFF state should be more
stable than the simplest LOFF state, and the simplest LOFF phase,
if it exists, will be replaced by a general LOFF phase. In this
sense, while the details of our phase diagrams obtained in the
following will be changed, the qualitative phase structure of the
system will remain, when a general LOFF state is included.

After a phase transformation for the fermion fields,
$\chi_\sigma=e^{i{\bf q}\cdot{\bf x}}\psi_\sigma$, the
thermodynamic potential in mean field approximation can be
evaluated as a summation of quasi-particles\cite{he2}
\begin{eqnarray}
\label{omega}
\Omega&=&-\frac{m\Delta^2}{4\pi
a_s}-\Delta^2\int\frac{d^3{\bf p}}{(2\pi)^3}\left(\frac{1}{E_p+\xi_p}-\frac{1}{2\epsilon_p}\right)\\
&&-\frac{1}{\beta}\int\frac{d^3{\bf
p}}{(2\pi)^3}\left[\ln\left(1+e^{-\beta
E_A}\right)+\ln\left(1+e^{-\beta E_B}\right)\right]\nonumber,
\end{eqnarray}
where $E_A$ and $E_B$ are the quasi-particle energies
\begin{eqnarray}
E_A&=&E_p+\delta\mu+{\bf p}\cdot{\bf q}/m,\nonumber\\
E_B&=&E_p-\delta\mu-{\bf p}\cdot{\bf q}/m
\end{eqnarray}
with $\xi_p=\left({\bf p}^2+{\bf q}^2\right)/(2m)-\mu$, $
E_p=\sqrt{\xi_p^2+\Delta^2}$, and average chemical potential $
\mu=\left(\mu_\uparrow+\mu_\downarrow\right)/2$ and chemical
potential difference
$\delta\mu=\left(\mu_\downarrow-\mu_\uparrow\right)/2$. Note that,
since the replacement of the bare coupling $g$ by the s-wave
scattering length $a_s$, the ultraviolet divergence in the first
momentum integration in (\ref{omega}) is removed, and there is no
need to introduce a momentum cutoff.

The thermodynamic potential obtained is a function of $T,
\mu_\uparrow$ and $\mu_\downarrow$ with $\Delta$ and ${\bf q}$
initially undetermined order parameter and pair momentum of the
superfluid. In the spirit of thermodynamics, the physical system
is described only by $T, \mu_\uparrow$ and $\mu_\downarrow$, and
$\Delta$ and ${\bf q}$ as functions of $T, \mu_\uparrow,
\mu_\downarrow$ are determined by the minimum thermodynamic
potential. The calculation of the first order derivatives of
$\Omega$ with respect to $\Delta$ and ${\bf q}$ gives the coupled
gap equations,
\begin{eqnarray}
\label{gap1}
-\frac{m\Delta}{4\pi a_s}&=& \Delta\int\frac{d^3{\bf
p}}{(2\pi)^3}\left[\frac{1-f(E_A)-f(E_B)}{2E_p}-\frac{1}{2\epsilon_p}\right],\nonumber\\
0&=& \int\frac{d^3{\bf p}}{(2\pi)^3}\Bigg[\frac{{\bf p}\cdot{\bf q}}{q}\left(f(E_B)-f(E_A)\right)\nonumber\\
&&-q\left(1-\frac{\xi_p}{E_p}\left(1-f(E_A)-f(E_B)\right)\right)\Bigg],
\end{eqnarray}
where we have chosen a suitable frame with the $z$ axis along the
direction of the pair momentum, ${\bf q}=(0,0,q)$, and $f(x)$ is
the Fermi-Dirac distribution function. It is easy to see that
$q=0$ is a trivial solution of the gap equations, which
corresponds to the homogeneous and isotropic phase. For a system
with fixed numbers of species, the chemical potentials
$\mu_\uparrow$ and $\mu_\downarrow$ in (\ref{gap1}) are solved
from the fermion number densities $n_\uparrow$ and $n_\downarrow$
derived by the thermodynamic relations $n_\sigma=-\partial
\Omega/\partial \mu_\sigma$,
\begin{equation}
\label{num} n_\uparrow =\int{d^3{\bf p}\over (2\pi)^3}n_\uparrow(
p),\ \ \ n_\downarrow =\int{d^3{\bf p}\over
(2\pi)^3}n_\downarrow(p),
\end{equation}
where $n_\uparrow(p)$ and $n_\downarrow(p)$ are the occupation
numbers for the two species,
\begin{eqnarray}
n_\uparrow(p)&=&u_p^2f(E_A)
+v_p^2f(-E_B),\nonumber\\
n_\downarrow(p)&=&u_p^2f(E_B)+v_p^2f(-E_A)
\end{eqnarray}
with the coherent coefficients $u_p^2=\left(1+\xi_p/E_p\right)/2$
and $v_p^2=\left(1-\xi_p/E_p\right)/2$.

The gap equations (\ref{gap1}) together with the fermion numbers
(\ref{num}) determine the physical values of $\Delta$ and $q$ as
functions of $n_\uparrow$ and $n_\downarrow$. Generally, there
exist a homogeneous solution with $q=0$ and a LOFF solution with
$q\neq 0$. To determine which one is favored at fixed fermion
numbers, we should compare their free energies defined as
\begin{equation}
\label{relation}
{\cal F}(n_\sigma,T)=\Omega+\mu_\uparrow
n_\uparrow+\mu_\downarrow n_\downarrow,
\end{equation}
and the lower one corresponds to the ground state of the system.

To explicitly describe the asymmetry between the two species, we
use the total density $n=n_\uparrow+n_\downarrow$ and the number
asymmetry parameter
$\alpha=(n_\downarrow-n_\uparrow)/(n_\downarrow+n_\uparrow)$ as
variables instead of $n_\uparrow$ and $n_\downarrow$. Without loss
of generality, we suppose $\alpha \ge 0$ in the following. Like a
free Fermi gas, we introduce a Fermi momentum $p_F$ or a Fermi
energy $E_F=p_F^2/(2m)$ through $n=p_F^3/(3\pi^2)$. If we scale
all the variables with energy dimension by $E_F$ and the variables
with momentum dimension by $p_F$, the solution of the gap
equations depends only on three dimensionless variables, the
coupling $p_Fa_s$, population imbalance $\alpha$ and the scaled
temperature $T/E_F$. For the following numerical calculation in
the BCS region with $-1<p_Fa_s<0$ or $-\infty<1/(p_Fa_s)<-1$, we
take $p_Fa_s=-0.6$. This is a typical value for systems with BCS
pairing. For example, for the pairing between $^6$Li atoms in
states $|F=3/2,m_F=3/2\rangle$ and $|F=3/2,m_F=1/2\rangle$ with
scattering length $a_s=-2160a_B$ where $a_B$ is the Borh radius
and typical density $n=3.8\times10^{12}$cm$^{-3}$, we have
$p_Fa_s=-0.56$.

\section {Temperature Behavior OF The Homogeneous Phase}
\label{s3}
In this section we focus on the homogeneous phase. We analyze the
temperature behavior of the pairing gap, the superfluid density
and the number susceptibility, and show that there should be a
phase transition from homogeneous phase at higher temperature to
some inhomogeneous phase at low temperature.

\subsection {Order Parameter $\Delta$}
At zero temperature, the gap and the chemical potential difference
for an asymmetric system with $\alpha\neq 0$ satisfy the relation
$\delta\mu>\Delta$, which can be seen directly from the number
equations. This means that the homogeneous phase at $T=0$ must be
a gapless superfluid\cite{he} with the gapless branch $E_B$. In
the BCS region, we have a positive chemical potential $\mu$ which
is not far from the Fermi energy $E_F$, and the chemical potential
mismatch $\delta\mu$ is much less than $\mu$, the two gapless
nodes occur at momenta
$p_1=\sqrt{2m(\mu-\sqrt{\delta\mu^2-\Delta^2})}$ and
$p_2=\sqrt{2m(\mu+\sqrt{\delta\mu^2-\Delta^2})}$. The behavior of
$\Delta$ is quite different from the one for symmetric BCS gap
$\Delta_0$ with the same coupling $p_Fa_s$. In weak coupling,
there is an analytical relation between them\cite{sarma},
\begin{equation}
\Delta(\delta\mu)=\sqrt{\Delta_0(2\delta\mu-\Delta_0)},\ \
\Delta_0/2<\delta\mu<\Delta_0
\end{equation}
in terms of the chemical potential mismatch or\cite{lombardor}
\begin{equation}
\label{dmu}\Delta(\alpha)=\Delta_0\sqrt{1-\alpha/\alpha_0}\Theta(\alpha_0-\alpha)
\end{equation}
in terms of the population imbalance,  where $\Theta(x)$ is a step
function and the critical imbalance $\alpha_0$ is given by
$\alpha_0=3\Delta_0/(4E_F)$. The solution $\delta\mu=\Delta_0$
corresponds to $\alpha=0$ and $\delta\mu=\Delta_0/2$ to
$\alpha=\alpha_0$.

\begin{figure}[!htb]
\begin{center}
\includegraphics[width=6cm]{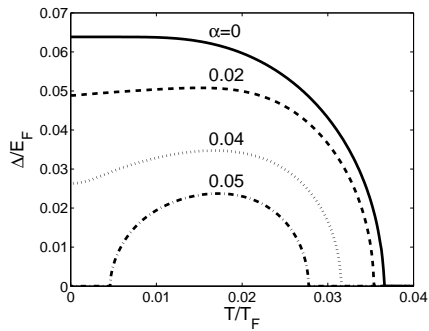}
\caption{The pairing gap $\Delta$, scaled by the Fermi energy
$E_F$, as a function of $T$, scaled by the Fermi temperature
$T_F=E_F/k_B$, for different values of number asymmetry $\alpha$.
\label{fig1}}
\end{center}
\end{figure}

For the symmetric system, it is well-known that the pairing gap
$\Delta_0$ will be suppressed by thermal motion at finite
temperature, as shown by the solid line marked with $\alpha=0$ in
Fig.\ref{fig1}. However, this monotonous temperature behavior is
not always true for the asymmetric system. The temperature effect
not only deforms and reduces the mismatched Fermi surfaces which
leads to the usual suppression of the gap, but also makes the
overlap region of the two species wider which favors the
condensate. The competition of the two opposite effects results in
a non-monotonous temperature behavior of the pairing gap. In
Fig.\ref{fig1}, we show the gap $\Delta$ as a function of
temperature for different values of population imbalance $\alpha$
and at the typical BCS coupling $p_F a_s=-0.6$. In the case with
weak imbalance, while the superfluidity starts still at $T=0$, it
firstly goes up with increasing temperature, then drops down, and
finally reaches zero at the critical temperature $T_c$. In the
case with strong imbalance, the superfluidity starts even at a
finite temperature $T_o>0$ and vanishes at $T_c$. The phase
transitions at the two critical temperatures $T_o$ and $T_c$ from
normal phase to HS is of second order, and the critical behavior
of the gap can be conventionally expressed as\cite{fetter}
\begin{eqnarray}
\label{cri} \Delta(T) &=& \phi T_o\left(T/T_o-1\right)^{1/2},\ \ \
T\rightarrow T_o^+,\nonumber\\
\Delta(T) &=& \varphi T_c\left(1-T/T_c\right)^{1/2},\ \ \
T\rightarrow T_c^-,
\end{eqnarray}
where $T_o, T_c$ and the dimensionless coefficients $\phi$ and
$\varphi$ all depend on the population imbalance $\alpha$. When
$\alpha$ exceeds the maximum value $\alpha_{HS}$ for the HS phase,
there is no more homogeneous superfluidity at any temperature. In
Fig.\ref{fig1} we have $\alpha_{HS}=0.059$.

\subsection {Superfluid Density $\rho_s$}
When the superfluid moves with a uniform velocity ${\bf v}_s$, the
condensate transforms like
$\langle\Phi\rangle\rightarrow\langle\Phi\rangle e^{2im{\bf
v}_s\cdot{\bf x}},
\langle\Phi^*\rangle\rightarrow\langle\Phi^*\rangle e^{-2im{\bf
v}_s\cdot{\bf x}}$, and the supercurrent ${\bf j}_s$ and the
superfluid density $\rho_s$ are defined via
\begin{equation}
\label{vel}
\Omega({\bf v}_s)=\Omega(0)+{\bf j}_s\cdot{\bf
v}_s+\frac{1}{2}\rho_s{\bf v}_s^2+\cdots.
\end{equation}
From the introduction (\ref{q}) for the pair momentum ${\bf q}$
and its gap equation, the supercurrent keeps zero,
\begin{equation}
{\bf j}_s={\partial\Omega\over\partial{\bf v}_s}\Big|_{{\bf
v}_s=0}=m{\partial\Omega\over\partial{\bf q}}\Big|_{{\bf q}=0}=0,
\end{equation}
and the superfluid density $\rho_s$ is directly related to the
pair momentum susceptibility
\begin{equation}
\label{kq1} \rho_s={\partial^2 \Omega\over\partial {\bf
v}^2_s}\Big|_{{\bf v}_s=0}=m^2 {\partial^2 \Omega\over\partial
{\bf q}^2}\Big|_{{\bf q}=0}.
\end{equation}
From the stability condition $\partial^2\Omega/\partial{\bf
q}^2\ge 0$ for the pair momentum ${\bf q}$, the superfluid density
controls the stability of the homogenous phase. When $\rho_s$ is
negative, the HS state is unstable. It is necessary to note that,
up to the square term, the expansions of the free energy ${\cal
F}$ in ${\bf v}_s$ and ${\bf q}$ are the same as the expansions of
$\Omega$. Therefore, we can safely use the definition (\ref{kq1})
to analyze the superfluid stability for the system with fixed
number densities\cite{wu2}.

Using the result in \cite{wu,he}, the superfluid density at finite
temperature can be explicitly expressed as
\begin{equation}
\label{kq3} \rho_s=mn+\int_0^\infty
dp\frac{p^4}{6\pi^2}\left[f^\prime(E_A)+f^\prime(E_B)\right]
\end{equation}
with the definition $f'(x)=df(x)/dx$. Note that the integration
term is always negative, and the asymmetry between the two species
may induce a negative superfluid density $\rho_s$. At zero
temperature and in weak coupling limit, $\rho_s$ can be
approximately expressed as\cite{he},
\begin{equation}
\label{kq5} \rho_s \approx
mn\left[1-\frac{\delta\mu\Theta(\delta\mu-\Delta)}{\sqrt{\delta\mu^2-\Delta^2}}\right].
\end{equation}
It is easy to see that in the Sarma phase with $\Delta<\delta\mu$,
$\rho_s$ becomes negative. This is the so called magnetic
instability, since it is directly related to the negative Meissner
mass squared $m_A^2=e^2\rho_s/m^2$\cite{he} if the fermions are
charged. This dynamical instability implies that the LOFF state
with finite pair momentum has lower free energy than the Sarma
state at zero temperature.

From the definition, the superfluid density vanishes at the
critical temperature $T_c$ for any imbalance
$0<\alpha<\alpha_{HS}$ and also at the other critical temperature
$T_o > 0$ when the imbalance is close to the maximum
$\alpha_{HS}$. In weak coupling case, the chemical potentials
$\mu_\uparrow$ and $\mu_\downarrow$ can be safely regarded as the
Fermi energies of the two species, and near the critical
temperatures $T_c$ and $T_o$ the superfluid density behaviors
as\cite{fetter}
\begin{eqnarray}
&& \rho_s = \rho_c\left(1-T/T_c\right)>0,\ \ \ T\rightarrow
T_c^-,\nonumber\\
&& \rho_s = \rho_o\left(1-T/T_o\right)<0,\ \ \ T\rightarrow T_o^+,
\end{eqnarray}
\begin{figure}[!htb]
\begin{center}
\includegraphics[width=6cm]{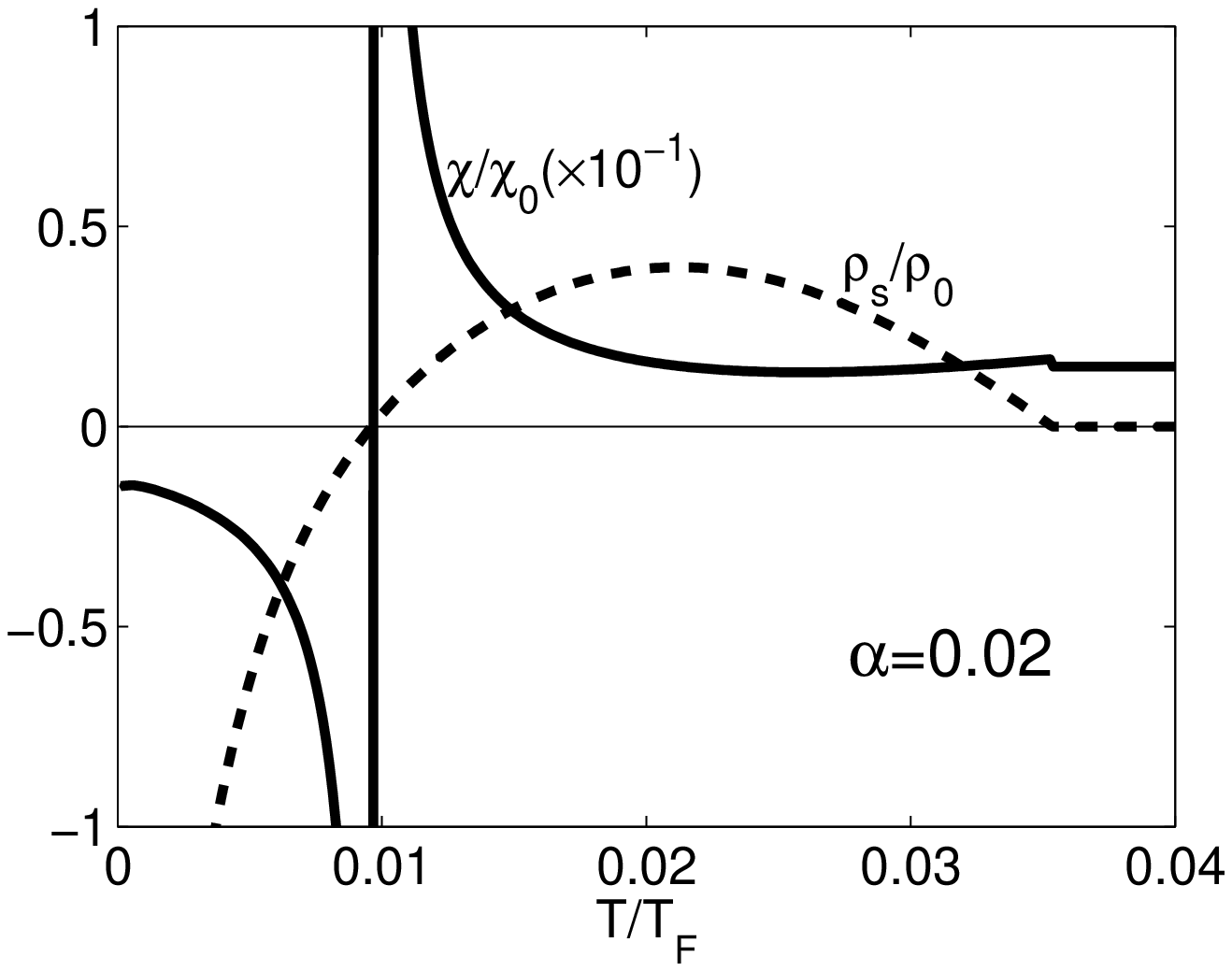}
\includegraphics[width=6cm]{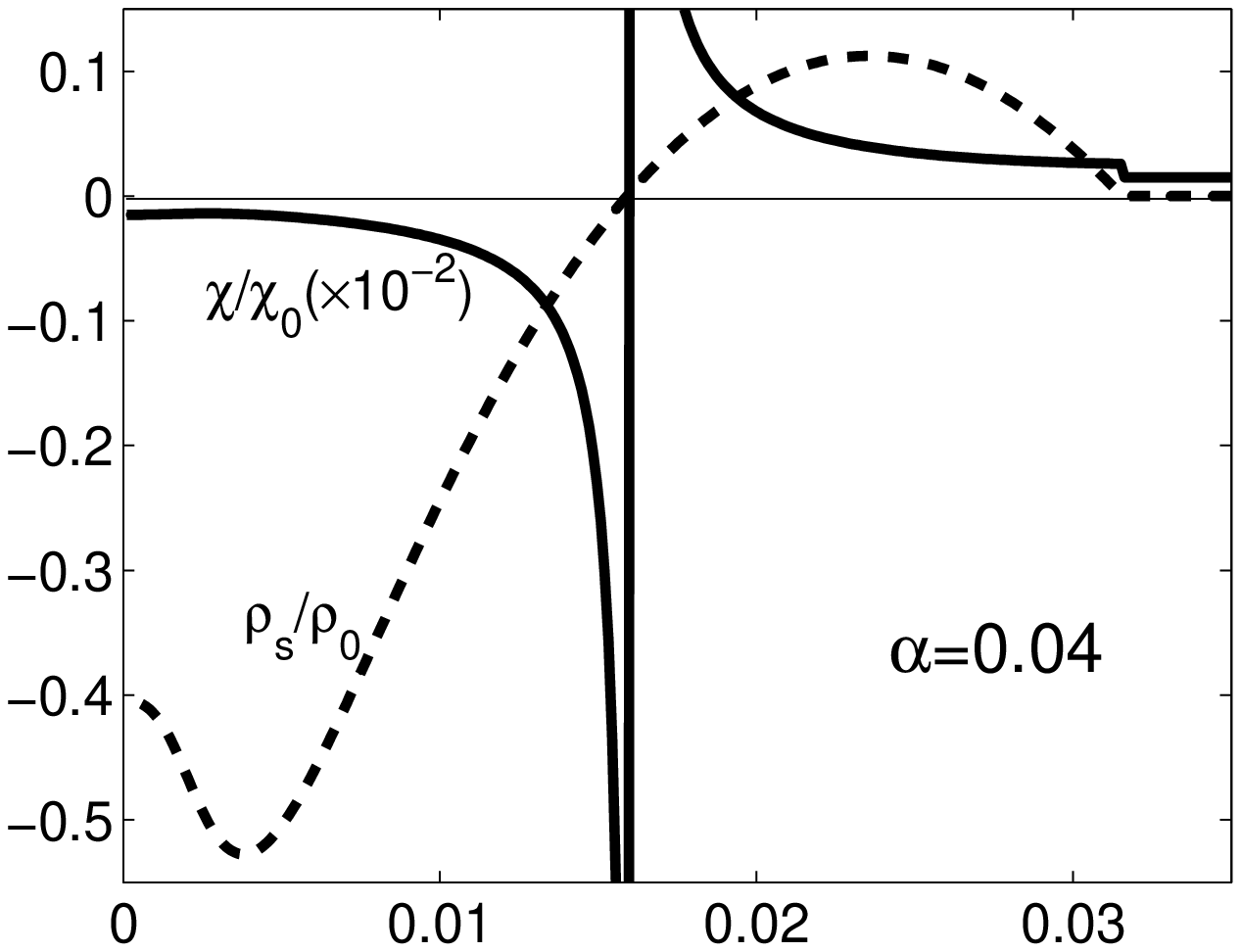}
\includegraphics[width=6cm]{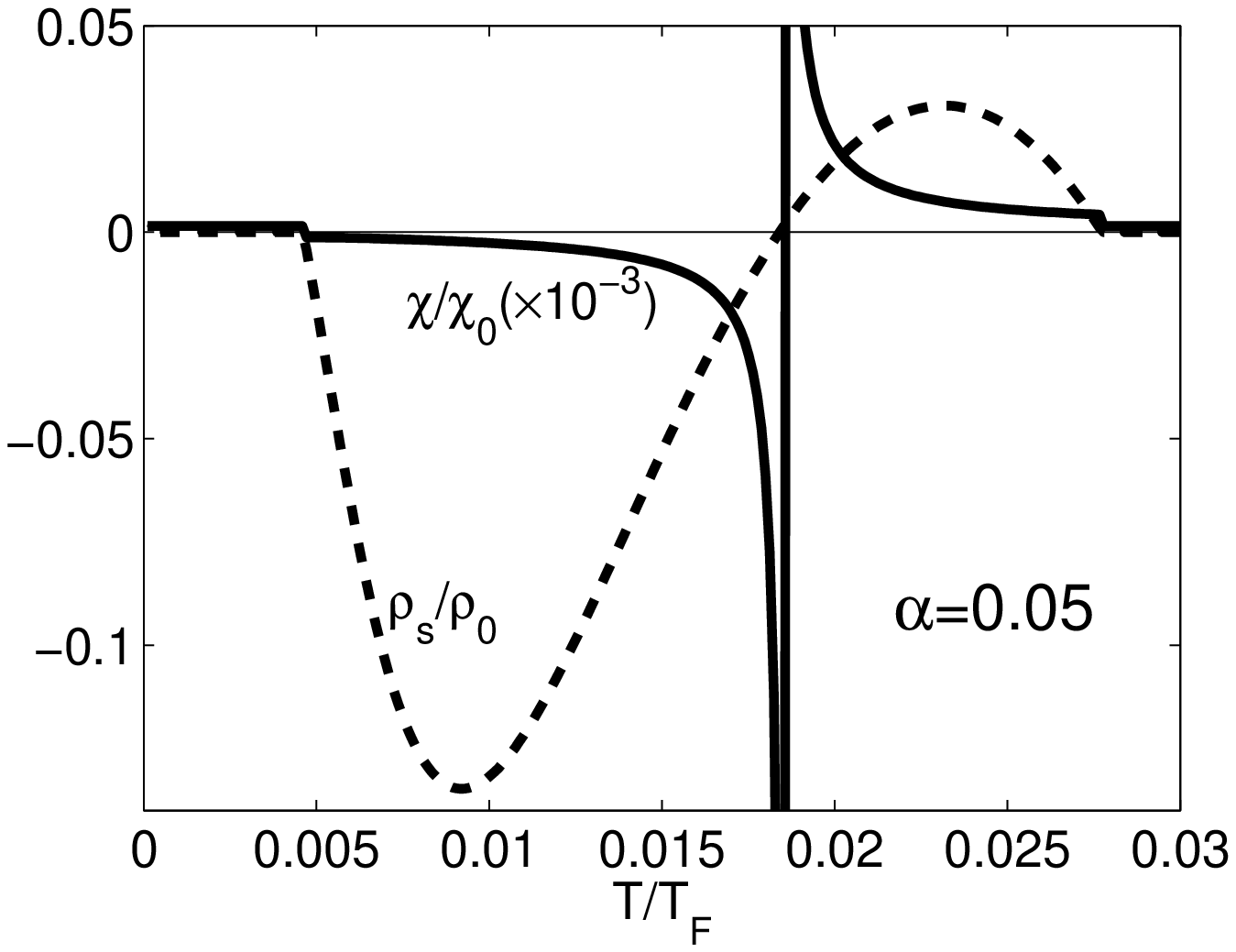}
\caption{The superfluid density $\rho_s$ (dashed lines), scaled by
$\rho_0=mn$, and number susceptibility $\chi$ (solid lines),
scaled by $\chi_0=n/E_F$, as functions of $T$, scaled by $T_F$,
for three values of number asymmetry in the BCS regime with
$p_Fa_s=-0.6$. \label{fig2}}
\end{center}
\end{figure}
where $\rho_c$ and $\rho_o$ depend only on the population
imbalance $\alpha$. Up to this point, we have shown that the
superfluid density is negative at low temperatures but should be
positive near the critical temperature. In Fig.\ref{fig2} we show
the superfluid density $\rho_s$ as a function of $T$ for different
values of population imbalance $\alpha$. For the symmetric system
with $\alpha=0$, $\rho_s$ is always positive in the whole
superfluidity region. For systems with weak imbalance, $\rho_s$ is
negative at low temperature, then changes sigh at an intermediate
temperature $T_1$, and keeps to be positive at high temperature.
When the imbalance is strong enough, $\rho_s$ is zero till the
starting temperature $T_o$ of the superfluidity, then becomes
negative first and turns to be positive at $T_1$. In
Fig.\ref{fig3}, we plot the two critical temperatures $T_o$ and
$T_c$ and the intermediate temperature $T_1$ as functions of
asymmetry parameter $\alpha$. The temperature $T_1$ separates the
stable HS phase with $\rho_s
>0$ at high temperature from the unstable HS phase with $\rho_s<0$
at low temperature. The negative superfluid density indicates that
the LOFF state may be energetically flavored at low temperature,
and there may be a phase transition from LOFF phase to HS phase at
$T=T_1$. We will come back to this point in the next section.
\begin{figure}[!htb]
\begin{center}
\includegraphics[width=6cm]{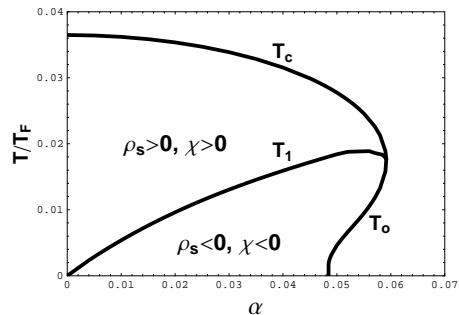}
\caption{The two critical temperatures $T_o$ and $T_c$ and the
intermediate temperature $T_1$ as functions of the asymmetry
parameter $\alpha$ in the BCS regime with $p_Fa_s=-0.6$. $\rho_s
> 0, \chi > 0$ and $\rho_s <0, \chi <0$ indicate, respectively,
the stable and unstable HS phases. \label{fig3}}
\end{center}
\end{figure}

\subsection {Number Susceptibility $\chi$}
The general stability condition for a two-component system against
changes in the densities of its components is described by the
total free energy of the system\cite{viverit,cohen}, $F=\int
d^3{\bf x}{\cal F}[n_\sigma({\bf x})]$. Considering its
fluctuations induced by small changes $\delta n_\sigma({\bf x})$,
the first-order variation $\delta F$ vanishes automatically due to
the number conservation, $\int d^3{\bf x}\delta n_\sigma({\bf
x})=0$, and the second-order variation $\delta^2F$ is given by the
quadratic form
\begin{equation}
\delta^2F=\frac{1}{2}\int d^3{\bf
x}\sum_{\sigma,\sigma'=\uparrow,\downarrow}\frac{\partial^2{\cal
F}}{\partial n_\sigma\partial n_{\sigma'}}\delta n_\sigma\delta
n_{\sigma'}.
\end{equation}
Therefore, to achieve a stable homogeneous phase, the $2\times 2$
matrix $\partial^2{\cal F}/\partial n_\sigma\partial n_{\sigma'}$
should be positively definite, namely, it has only positive eigen
values. This matrix is hard to calculate since it is not easy to
express explicitly the free energy as a function of densities.
However, from the relation (\ref{relation}) between the free
energy ${\cal F}(n_\sigma)$ and thermodynamic potential
$\Omega(\mu_\sigma)$, it is easy to check that the stability
condition to have positively definite matrix $\partial^2{\cal
F}/\partial n_\sigma\partial n_{\sigma'}$ is equivalent to the
condition to have positively definite matrix
$-\partial^2\Omega/\partial \mu_\sigma\partial \mu_{\sigma'}$.

For systems without mass difference between the two species, the
condition to have positive eigenvalues of
$-\partial^2\Omega/\partial \mu_\sigma\partial \mu_{\sigma'}$ can
be reduced to the constraint\cite{pao} that the imbalance number
susceptibility
$\chi=-(\partial^2\Omega/\partial\delta\mu^2)_\mu=\left(\partial\delta
n/\partial\delta\mu\right)_{\mu}$ should be positive. For
$\chi<0$, the density difference $\delta
n=n_\downarrow-n_\uparrow$ increases with decreasing chemical
potential difference $\delta\mu$, which is certainly unphysical
and means instability of the superfluid against the phase
separation (PS). Employing the gap equation which determines the
condensate as a function of chemical potentials,
$\Delta=\Delta(\mu_\sigma)$, we can express the number
susceptibility $\chi$ as a direct and indirect parts,
\begin{equation}
\chi=\left(\frac{\partial\delta
n}{\partial\delta\mu}\right)_{\mu,\Delta}+\left(\frac{\partial\delta
n}{\partial\Delta}\right)_{\mu,\delta\mu}\left(\frac{\partial\Delta}{\partial\delta\mu}\right)_\mu.
\end{equation}

From the expression
\begin{eqnarray}
\label{deltan}
\delta n=\int{d^3{\bf p}\over
(2\pi)^3}\left[f(E_B)-f(E_A)\right]
\end{eqnarray}
which leads to
\begin{eqnarray}
\label{ddeltan}
\left(\frac{\partial\delta n}{\partial\delta\mu}\right)_{\Delta,\mu}&=&-\int{d^3{\bf p}\over (2\pi)^3}
\left[f^\prime(E_A)+f^\prime(E_B)\right],\nonumber\\
\left(\frac{\partial\delta
n}{\partial\Delta}\right)_{\mu,\delta\mu}&=&\int{d^3{\bf p}\over
(2\pi)^3}\frac{\Delta}{E_p}\left[f^\prime(E_B)-f^\prime(E_A)\right],
\end{eqnarray}
and the gap equation
$\left(\partial\Omega/\partial\Delta\right)_{\mu,\delta\mu}=0$
which results in
\begin{equation}
\left[\frac{\partial}{\partial\delta\mu}\left(\frac{\partial
\Omega}{\partial\Delta}\right)_{\mu,\delta\mu}\right]_{\mu,\Delta}+\left(\frac{\partial^2
\Omega}{\partial
\Delta^2}\right)_{\mu,\delta\mu}\left(\frac{\partial\Delta}{\partial\delta\mu}\right)_{\mu}=0,
\end{equation}
namely,
\begin{equation}
\left(\frac{\partial\Delta}{\partial\delta\mu}\right)_{\mu}=\left(\frac{\partial
\delta
n}{\partial\Delta}\right)_{\mu,\delta\mu}\left(\frac{\partial^2
\Omega}{\partial \Delta^2}\right)_{\mu,\delta\mu}^{-1},
\end{equation}
we have
\begin{equation}
\label{chi}
\chi=\left(\frac{\partial\delta
n}{\partial\delta\mu}\right)_{\mu,\Delta}+\left(\frac{\partial
\delta
n}{\partial\Delta}\right)^2_{\mu,\delta\mu}\left(\frac{\partial^2
\Omega}{\partial \Delta^2}\right)_{\mu,\delta\mu}^{-1}.
\end{equation}
Since $(\partial\delta n/\partial\delta\mu)_{\mu,\Delta}$ is
always positive, the stability condition is controlled by the gap
susceptibility
$\kappa=(\partial^2\Omega/\partial\Delta^2)_{\mu,\delta\mu}$ which
determines if the solution of the gap equation is the minimum of
the thermodynamic potential.

We now consider the relation between the gap susceptibility and
the superfluid density. It is easy to explicitly write down the
gap susceptibility,
\begin{eqnarray}
\label{domega}
\kappa&=&\int{d^3{\bf p}\over
(2\pi)^3}\frac{\Delta^2}{E_p^2}\bigg[\frac{1-f(E_A)-f(E_B)}{E_p}\nonumber\\
&&+\left(f^\prime(E_A)+f^\prime(E_B)\right)\bigg]\nonumber\\
&&+\frac{2}{g}-\int{d^3{\bf p}\over
(2\pi)^3}\frac{1-f(E_A)-f(E_B)}{E_p},
\end{eqnarray}
where the last line vanishes automatically due to the gap equation
in the superfluid phase. In weak coupling limit, the number
susceptibility and gap susceptibility in the superfluid phase at
zero temperature can be evaluated as
\begin{equation}
\chi=-\frac{3n}{2E_F},\ \ \ \kappa=\frac{mp_F}{\pi^2}\left[1-
\frac{\delta\mu\theta(\delta\mu-\Delta)}{\sqrt{\delta\mu^2-\Delta^2}}\right]
\end{equation}
which are negative at any imbalance $\alpha$. On the other hand,
applying partial integration to the number equations, the
superfluid density can be expressed as
\begin{eqnarray}
\rho_s&=&\int{d^3{\bf p}\over (2\pi)^3}\frac{{\bf
p}^2}{3}\frac{\Delta^2}{E_p^2}\bigg[\frac{1-f(E_A)-f(E_B)}{E_p}\nonumber\\
&&+\left(f^\prime(E_A)+f^\prime(E_B)\right)\bigg].
\end{eqnarray}
Comparing $\kappa$ and $\rho_s$ in the superfluid phase, the only
difference is the factor of ${\bf p}^2/3$ in the integrand
function of $\rho_s$. In the BCS region, the integration is
dominated by a narrow momentum window around the average Fermi
momentum $p_0=\sqrt{2m\mu}$, and we have
\begin{equation}
\label{krhos} \rho_s\simeq \frac{p_0^2}{3}\kappa.
\end{equation}
From this proportional relation between $\kappa$ and $\rho_s$ and
note that $\kappa$ is the denominator of the second term of
(\ref{chi}), while the superfluid density and number
susceptibility have different amplitudes, they change sign almost
at the same position. We have checked numerically that such an
approximation holds well in the BCS region
$-\infty<1/(p_Fa_s)<-1$. When $\rho_s$ changes from $\rho_s>0$ to
$\rho_s<0$ at the turning temperature $T_1$, $\chi$ changes from
positive infinity to negative infinity approximately at the same
temperature. Therefore, when $\rho_s$ is not too large, namely
when the second term of (\ref{chi}) is guaranteed to dominate
$\chi$, $\rho_s$ and $\chi$ have almost the same positive and
negative regions in the BCS limit. This temperature behavior of
superfluid density and number susceptibility is clearly shown in
Fig.\ref{fig2} where we presented $\rho_s$ and $\chi$ as functions
of temperature. From the stability analysis at different
population imbalance $\alpha$, we plot in Fig.\ref{fig3} the phase
diagram of the homogeneous superfluid in the $T-\alpha$ plane for
the coupling $p_Fa_s=-0.6$. The temperature $T_1(\alpha)$
separates the stable HS phase at high temperature where $\chi$ and
$\rho_s$ are positive from the unstable HS phase at low
temperature where $\chi$ and $\rho_s$ are negative. Because of the
divergence of $\chi$ at the turning point $T_1$, the possible
phase transition from PS to HS must be of first order.

From the above analytic and numerical calculations in the BCS
region, the conditions to guarantee a stable homogeneous
superfluid described by the superfluid density, number
susceptibility and gap susceptibility are equivalent. We can show
that this is also true for unequal mass system. For instance, the
breached pairing state proposed in \cite{forbes} which is the
minimum of the thermodynamic potential should also be stable
against the LOFF state and PS state. Certainly, when the coupling
is strong enough, the difference among these stable conditions
will become remarkable. Due to the factor of ${\bf p}^2/3$, the
superfluid density is easier to be positive in strong coupling
region than the number susceptibility, this is the reason why the
authors in \cite{pao} found that the positive superfluid density
is a weaker condition than the positive number susceptibility. At
finite temperature, we found that for a strong enough coupling,
the turning temperature $T_1$ for the superfluid density $\rho_s$
is remarkably different from the one for the number susceptibility
$\chi$. Generally, the superfluid density $\rho_s$ becomes
positive at a lower temperature, namely, there exists a large
region where $\rho_s$ is positive but $\chi$ is negative. However,
when the coupling is strong enough, the naive mean field treatment
is not valid at finite temperature and we should employ some
proper finite temperature theory for BCS-BEC crossover\cite{chen}.

In the BCS region considered in this paper, from Figs.\ref{fig2}
and \ref{fig3}, the LOFF and PS are both possible phases of the
superfluid at low temperature $T<T_1$, the question left is to
determine which of them is stable.

\section {LOFF Phase vs Homogeneous Phase}
\label{s4}
We now take the LOFF state into account. Firstly we consider the
LOFF state with $\alpha < \alpha_{HS}$ where the HS phase can
survive. Before numerically solving the coupled gap equations
(\ref{gap1}) for $\Delta$ and $q$ together with the number
equations (\ref{num}) for $n_\uparrow$ and $n_\downarrow$, we
discuss analytically the pair momentum in the temperature region
close to the turning point $T_1$. The first order derivative of
the thermodynamic potential with respect to the pair momentum can
be written as
\begin{equation}
\frac{\partial \Omega}{\partial q}=qW(q)
\end{equation}
which gives the gap equation for the pair momentum, $qW(q)=0$. The
homogeneous and isotropic state corresponds to the trivial
solution $q=0$, and the solution corresponding to the LOFF state
is given by $W(q)=0$. From the second derivative,
\begin{equation}
\frac{\partial^2\Omega}{\partial q^2}=W(q)+qW^\prime(q)
\end{equation}
and the relation between the superfluid density and the pair
momentum susceptibility demonstrated in Section \ref{s3}, we have
\begin{equation}
\rho_s=m^2W(0)
\end{equation}
which means
\begin{equation}
W(0)=0
\end{equation}
at $T_1$ where $\rho_s$ changes sign continuously. Providing the
LOFF solution is unique (we have checked this numerically at least
in the BCS region), the LOFF momentum $q$ must vanish at $T_1$,
and in the temperature region below and close to $T_1$, $q$ is
very small and we have approximately the relation between the free
energies for LOFF state and HS state,
\begin{equation}
{\cal F}_{LOFF}={\cal F}_{HS}+\frac{\rho_s}{2m^2}q^2+O(q^4).
\end{equation}
Therefore, in the unstable HS region with $\rho_s<0$ at $T\le
T_1$, the LOFF state has lower free energy than the HS state, and
the critical behavior of the LOFF momentum is
\begin{equation}
\label{cq} q \sim \left(1-T/T_1\right)^{1/2},\ \ \ T\rightarrow
T_1^-.
\end{equation}
\begin{figure}[!htb]
\begin{center}
\includegraphics[width=6cm]{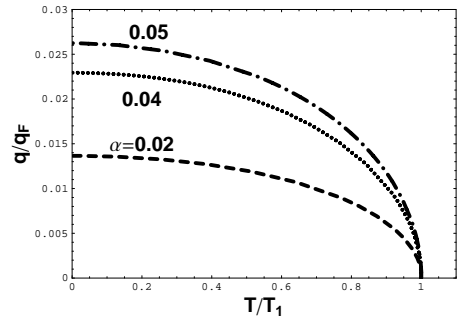}
\caption{The LOFF momentum $q$, scaled by the Fermi momentum
$p_F$, as a function of $T$, scaled by the intermediate
temperature $T_1$, for several values of number asymmetry
$\alpha$. \label{fig4}}
\end{center}
\end{figure}

\begin{figure}[!htb]
\begin{center}
\includegraphics[width=6cm]{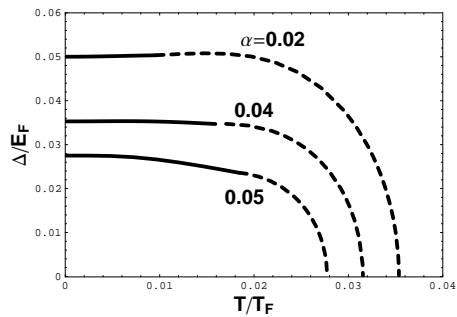}
\caption{The pairing gap $\Delta$, scaled by $E_F$, as a function
of $T$, scaled by $T_F$, for several values of number asymmetry
$\alpha$. The solid lines are for LOFF state and the dashed lines
for BP state. \label{fig5}}
\end{center}
\end{figure}

In Fig.\ref{fig4}, we show the LOFF momentum $q$ scaled by the
Fermi momentum $p_F$ as a function of temperature scaled by the
intermediate temperature $T_1$ for several values of number
asymmetry $\alpha$. At any $\alpha$, the momentum drops down from
the maximum at $T=0$ to zero at $T=T_1$. Note that $T_1$ is
$\alpha$ dependent, it increases with increasing asymmetry. In
Fig.\ref{fig5}, we demonstrate the pairing gap $\Delta$ for the
LOFF state (solid lines) and HS state (dashed lines). The LOFF
state survives only in the region $0<T<T_1$, and the stable HS
state exists in the region $T_1<T<T_c$. The two meet at the
intermediate temperature $T_1$ with the continuity
$\Delta_{LOFF}(T_1)=\Delta_{HS}(T_1)$. Different from the pair
momentum which increases with increasing asymmetry, the gap
parameter decreases with increasing asymmetry. It is necessary to
point out that the LOFF state can survive not only in the unstable
HS region, but also in the region where the HS phase disappears.
For instance, for $\alpha=0.05$, the HS phase disappears in the
region $0<T<T_o$, but LOFF state can survive there.
\begin{figure}[!htb]
\begin{center}
\includegraphics[width=6cm]{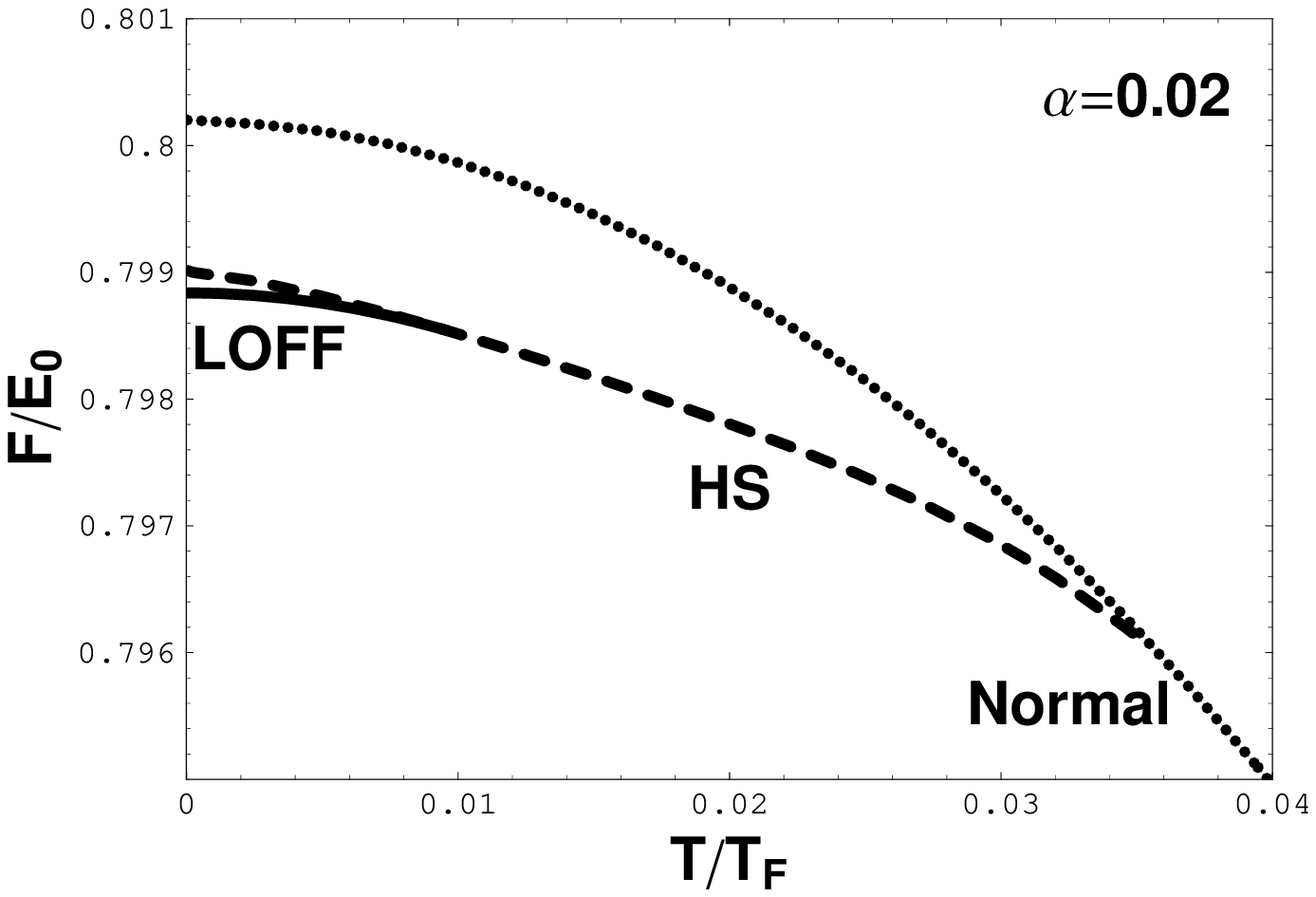}
\includegraphics[width=6cm]{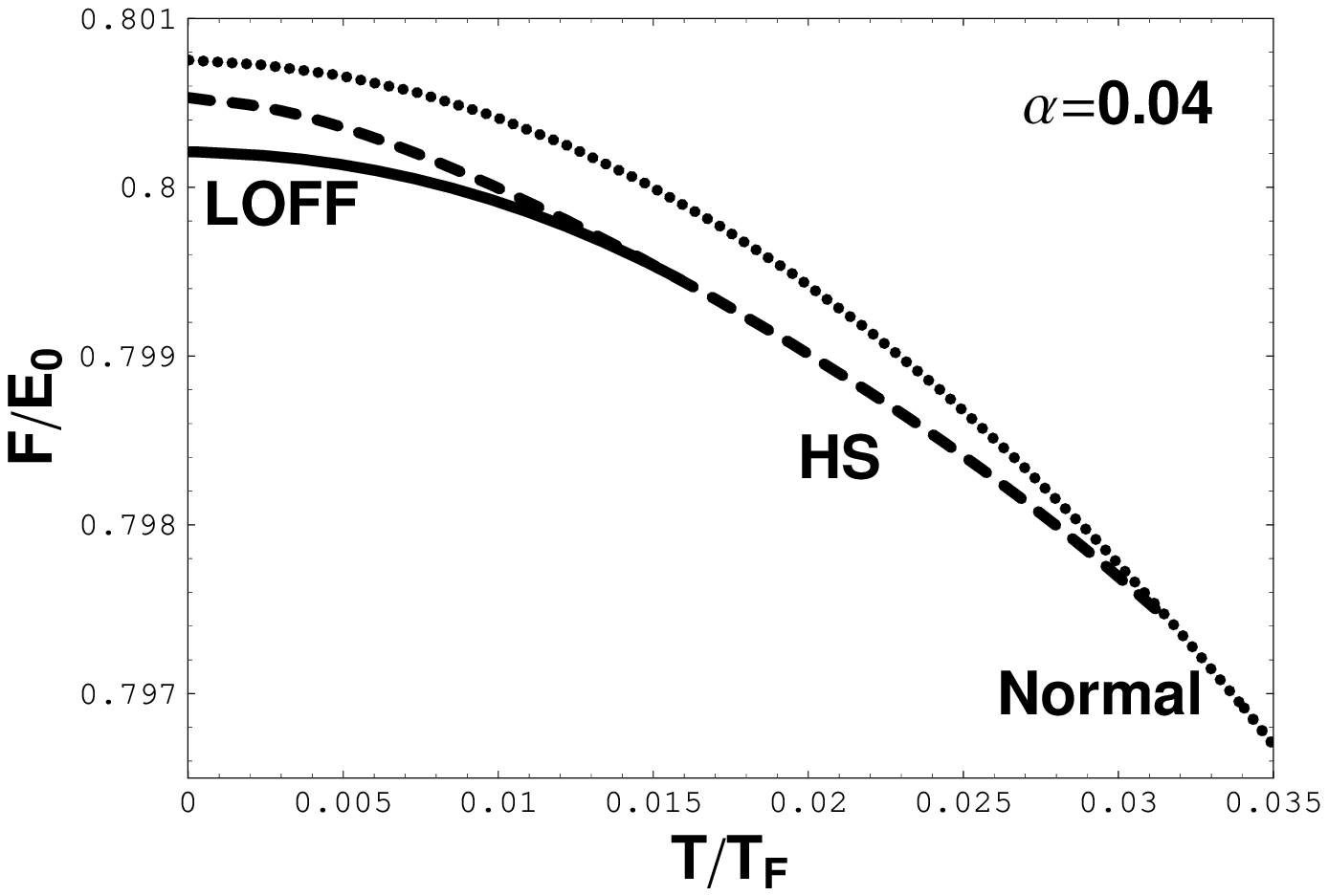}
\includegraphics[width=6cm]{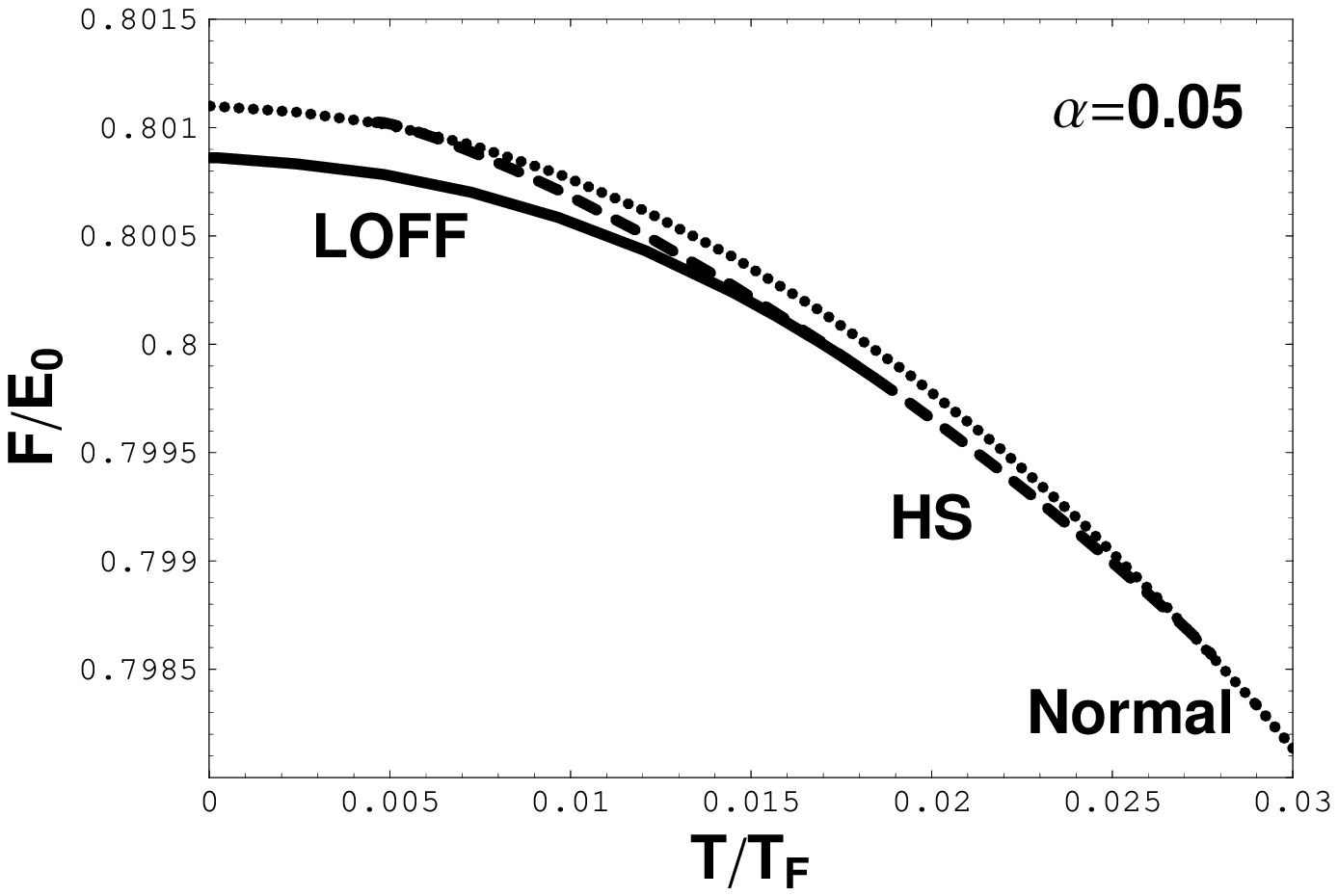}
\caption{The free energy ${\cal F}$, scaled by
$E_0=p_F^5/(8\pi^2m)$, as a function of $T$, scaled by $T_F$, for
several values of number asymmetry $\alpha$. The solid, dashed and
dotted lines correspond, respectively to the LOFF, HS and normal
states. \label{fig6}}
\end{center}
\end{figure}

In Fig.\ref{fig6} we compare the free energies for the LOFF, HS
and normal state for different values of asymmetry $\alpha$. It is
clear that LOFF is more stable than the other two in the region
$0<T<T_1$, and HS is stable than the normal state in the region
$T_1<T<T_c$. The phase transition from HS to LOFF happens at the
intermediate temperature $T_1$, and the order parameter
characterizing the spontaneous rotational symmetry breaking is the
LOFF momentum $q$. Since $q$ drops down monotonously to zero when
$T$ approaches to $T_1$, the phase transition is of second order.
From the above result, the strange superfluidity window in the
temperature region $T_o<T<T_c$ obtained in\cite{chen,sedrakian3}
disappears, when the LOFF state is taken into account. The
superfluid can exist in the whole temperature region $0<T<T_c$
with different phase at different temperature. When the
temperature is higher than the critical value $T_c$, the strong
thermal motion suppresses any superfluidity, and the system is in
normal Fermi gas. From the continuous temperature behavior
(\ref{cri}) of the HS gap, the phase transition at $T_c$ is of
second order.
\begin{figure}[!htb]
\begin{center}
\includegraphics[width=6cm]{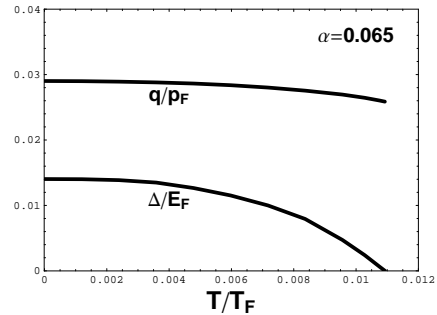}
\caption{The LOFF pairing gap $\Delta$, scaled by $E_F$, and the
momentum, scaled by $p_F$, as functions of $T$, scaled by $T_F$,
at $\alpha=0.065>\alpha_{HS}$. \label{fig7}}
\end{center}
\end{figure}
\begin{figure}[!htb]
\begin{center}
\includegraphics[width=6cm]{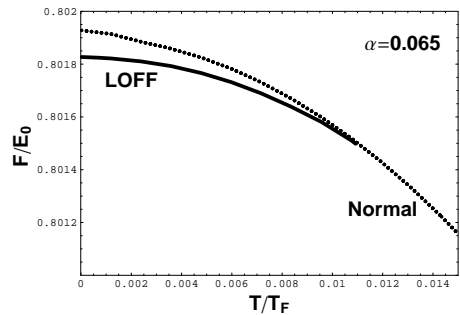}
\caption{The free energy ${\cal F}$, scaled by $E_0$, as a
function of $T$, scaled by $T_F$, at $\alpha=0.065>\alpha_{HS}$.
The solid and dotted lines are, respectively, for the LOFF and
normal phases. \label{fig8}}
\end{center}
\end{figure}

In the above discussion we have restricted ourself in the
asymmetry region $\alpha<\alpha_{HS}$ where the HS phase can
survive. What is about the case with $\alpha > \alpha_{HS}$, and
can the LOFF further survive in high asymmetric systems where
there is no room for the HS? In Fig.\ref{fig7} we show the LOFF
momentum $q$ and gap $\Delta$ as functions of temperature for
$\alpha=0.065>\alpha_{HS}$. Both drops down monotonously with
increasing temperature, and the gap approaches to zero
continuously. To determine the stable state, we compare the free
energies for the LOFF and normal phase in Fig.\ref{fig8}. It is
easy to see that the LOFF is stable at low temperature and the
normal phase becomes the only possible state at high temperature,
and the phase transition from the normal phase to LOFF is of
second order. The fact that the LOFF can survive at higher density
asymmetry where the HS disappears is consistent with the
conclusion at zero temperature\cite{he2}. Certainly, when the
asymmetry is beyond the maximum value $\alpha_{LOFF}$ for the LOFF
state, the system is in the normal phase at any temperature.

\section {Stability of LOFF Phase}
\label{s5}
In Section \ref{s4} we neglected the possibility of the PS phase
and focused only on the LOFF phase. Since both PS and LOFF phases
can be the ground state in the temperature region $0<T<T_1$, we
analyze in this section the stability of the known LOFF phase
against the PS. To this end, we study again the imbalance number
susceptibility $\chi=\left(\partial\delta
n/\partial\delta\mu\right)_{\mu}$ for the LOFF state,
\begin{eqnarray}
\chi&=&\left(\frac{\partial\delta
n}{\partial\delta\mu}\right)_{\mu,\Delta,q}+\left(\frac{\partial\delta
n}{\partial\Delta}\right)_{\mu,\delta\mu,q}\left(\frac{\partial\Delta}{\partial\delta\mu}\right)_{\mu,q}\nonumber\\
&&+\left(\frac{\partial\delta n}{\partial
q}\right)_{\mu,\delta\mu,\Delta}\left(\frac{\partial
q}{\partial\delta\mu}\right)_{\mu,\Delta}.
\end{eqnarray}
To simplify the expression, we ignore in the following the
subscript notes and this will not make any confusion.

Similar to the calculation for the HS phase, from the number
difference $\delta n$ (\ref{deltan}), we obtain the same
expression (\ref{ddeltan}) for the derivatives $\partial\delta
n/\partial\delta\mu$ and $\partial\delta n/\partial\Delta$ and the
derivative
\begin{equation}
\frac{\partial\delta n}{\partial q}=-\frac{1}{m}\int{d^3{\bf
p}\over (2\pi)^3}\frac{{\bf p}\cdot{\bf
q}}{q}\left[f^\prime(E_A)+f^\prime(E_B)\right].
\end{equation}
With the derivatives $\partial\Delta/\partial\delta\mu$ and
$\partial q/\partial\delta\mu$ evaluated from the coupled gap
equations for $\Delta$ and $q$,
\begin{eqnarray}
\frac{\partial^2 \Omega}{\partial \Delta
\partial \delta\mu}+\frac{\partial^2 \Omega
}{\partial
\Delta^2}\frac{\partial\Delta}{\partial\delta\mu}+\frac{\partial^2
\Omega}{\partial \Delta
\partial q}\frac{\partial q}{\partial\delta\mu}&=&0,\nonumber\\
\frac{\partial^2 \Omega}{\partial q
\partial \delta\mu}+\frac{\partial^2 \Omega
}{\partial q \partial
\Delta}\frac{\partial\Delta}{\partial\delta\mu}+\frac{\partial^2
\Omega}{\partial q^2}\frac{\partial q}{\partial\delta\mu}&=&0,
\end{eqnarray}
we have
\begin{equation}
\chi=\frac{\partial\delta
n}{\partial\delta\mu}+\frac{\left(\frac{\partial\delta
n}{\partial\Delta}\right)^2\frac{\partial^2 \Omega}{\partial
q^2}+\left(\frac{\partial\delta n}{\partial
q}\right)^2\frac{\partial^2 \Omega}{\partial
\Delta^2}-2\frac{\partial\delta
n}{\partial\Delta}\frac{\partial\delta n}{\partial
q}\frac{\partial^2 \Omega }{\partial q \partial \Delta}}{\det
{\cal M}},
\end{equation}
where ${\cal M}$ is the stability matrix for the LOFF state
defined as
\begin{equation}
{\cal M}=\left(\begin{array}{cc} \frac{\partial^2 \Omega}{\partial
\Delta^2}&\frac{\partial^2 \Omega }{\partial \Delta \partial q}
\\ \frac{\partial^2 \Omega }{\partial q \partial \Delta}&\frac{\partial^2 \Omega }{\partial
q^2}\end{array}\right)
\end{equation}
with the gap susceptibility $\partial^2\Omega/\partial\Delta^2$
given in (\ref{domega}) and the other two elements,
\begin{eqnarray}
\frac{\partial^2 \Omega}{\partial q^2}&=&\frac{n}{m}+{1\over
m^2}\int{d^3{\bf
p}\over(2\pi)^3}\bigg[\frac{({\bf p}\cdot{\bf q})^2}{q^2}\left(f^\prime(E_A)+f^\prime(E_B)\right)\nonumber\\
&&+2{\bf p}\cdot{\bf q}\frac{\xi_p}{E_p}\left(f^\prime(E_A)-f^\prime(E_B)\right)\nonumber\\
&&+q^2\frac{\xi_p^2}{E_p^2}\left(f^\prime(E_A)+f^\prime(E_B)\right)\nonumber\\
&&-q^2\frac{\Delta^2}{E_p^3}\left(1-f(E_A)-f(E_B)\right)\bigg].\nonumber\\
\frac{\partial^2 \Omega}{\partial \Delta \partial q}&=&{1\over
m}\int{d^3{\bf p}\over
(2\pi)^3}\bigg[\frac{{\bf p}\cdot{\bf q}}{q}\frac{\Delta}{E_p}\left(f^\prime(E_A)-f^\prime(E_B)\right)\nonumber\\
&&+q\frac{\Delta\xi_p}{E_p^2}\left(f^\prime(E_A)+f^\prime(E_B)\right)\nonumber\\
&&+q\frac{\Delta\xi_p}{E_p^3}\left(1-f(E_A)-f(E_B)\right)\bigg].
\end{eqnarray}

\begin{figure}[!htb]
\begin{center}
\includegraphics[width=6cm]{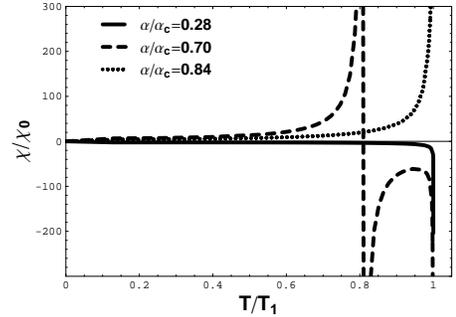}
\caption{The number susceptibility $\chi$ for the LOFF state,
scaled by its value $\chi_0$ in the symmetric system with $\alpha
= 0$, as a function of temperature $T$, scaled by $T_1$, for
several values of number asymmetry $\alpha$. \label{fig9}}
\end{center}
\end{figure}

The number susceptibility for the LOFF state is illustrated in
Fig.\ref{fig9} as a function of temperature in the region
$0<T/T_1<1$. For $\alpha<0.53 \alpha_{LOFF}$, $\chi$ is always
negative in the whole region, which means unstable LOFF state.
This instability of the LOFF state is consistent with the
conclusion obtained in \cite{hu,he2} at zero temperature. For
$0.53\alpha_{LOFF}<\alpha<\alpha_{HS}$, $\chi$ is positive at low
temperature, then changes sign at the turning temperature $T_2$,
and keeps to be negative at higher temperature between $T_2$ and
$T_1$. Similar to the number susceptibility in homogeneous case,
it is divergent at the turning temperature $T_2$. When $\alpha$ is
larger than the maximum asymmetry $\alpha_{HS}$ for the HS phase
but less the maximum value $\alpha_{LOFF}$ for the LOFF phase,
$\chi$ is always positive in the whole temperature region.

\begin{figure}[!htb]
\begin{center}
\includegraphics[width=6cm]{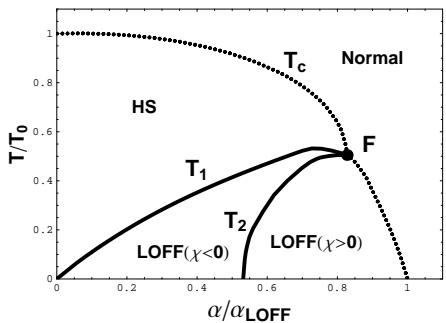}
\caption{The phase diagram in $T-\alpha$ plane. The solid lines
$T_1$ and $T_2$ denote, respectively, the first order phase
transitions from HS to PS and from LOFF to PS, the dashed line
means the second order phase transition from HS or LOFF to normal
phase, and F is the tetracritical point. \label{fig10}}
\end{center}
\end{figure}
From the analytic and numerical calculations for the HS, LOFF, PS
and normal state in this section and Sections \ref{s2}, \ref{s3}
and \ref{s4}, the ground states for the asymmetric system at
different temperature and asymmetry are summarized in the phase
diagram in $T-\alpha$ plane in Fig.\ref{fig10}. The dashed line
separates the superfluid state at low temperature and asymmetry
from the normal state at high temperature and asymmetry. Inside
the superfluid state, there are three phases, the homogeneous
superfluid above the temperature $T_1$ marked by HS, the stable
LOFF phase on the right hand side of the temperature line $T_2$
marked by LOFF $(\chi > 0)$, and the unstable LOFF phase in
between the temperature lines $T_1$ and $T_2$ marked by LOFF
$(\chi < 0)$ which is probably the PS phase. Since the number
susceptibilities for HS and LOFF are both divergent at the
corresponding turning temperatures $T_1$ and $T_2$, the phase
transitions from LOFF to possible PS and from HS to PS are of
first order. On the other hand, the phase transitions from HS to
normal state and from LOFF to normal state are of second order.
The four phases, HS, PS, LOFF and normal state meet at the point F
located at $(T,\alpha) = (0.49 T_0,
\alpha_{HS}=0.83\alpha_{LOFF})$ where $T_0$ is the critical
temperature $T_c$ in the symmetric system with $\alpha=0$. The
phase diagram we obtained here for the system with fixed densities
is quite different from the one for the system with fixed chemical
potentials\cite{takada,casalbuoni} in $T-\delta\mu$ plane. There
is no room for the PS phase in the $T-\delta\mu$ plane, and the
window for the LOFF and PS states in Fig.\ref{fig10} is much
larger than the one for the LOFF state in $T-\delta\mu$ plane.
While there exists a phase transition line from LOFF to HS in
$T-\delta\mu$ plane, it disappears in the $T-\alpha$ plane.

We have checked in the frame of mean field approximation that the
topological structure of the phase diagram in $T-\alpha$ plane
does not change for a wide BCS region $-\infty<1/(p_Fa_s)<-1$. If
pair fluctuation is included, the second order phase boundary
between the normal phase and HS or LOFF phase will be shifted, but
the topological structure will remain unchanged. Our phase diagram
is similar to the one obtained in \cite{machida} based on the BdG
formalism in an atomic trap.

While our result that the phenomenon of intermediate temperature
superfluid shown in Fig.\ref{fig1} is washed out by introducing
inhomogeneous pairing state is obtained by considering the
simplest single plane wave LOFF state, the qualitative conclusion
may remain unchanged when we take into account a more complicated
LOFF state, since a general inhomogeneous state will produce a
deeper minimum of the free energy of the system. On the other
hand, we can restudy this problem by investigating the stability
of the normal phase against an inhomogeneous
fluctuation\cite{larkin,fulde,ren}. Let us consider a static but
inhomogeneous pair fluctuation $\Phi({\bf x})$ for the normal
state in Fig.\ref{fig3}. The effective action of the system can be
expressed as a series of the fluctuation,
\begin{equation}
S_\text{eff}=S_0+S_2+O(\Phi^4),
\end{equation}
where $S_0$ is the action of the normal Fermi gas. The quadratic
term $S_2$ reads
\begin{equation}
S_2=\frac{1}{2}\int{d^3{\bf k}\over (2\pi)^3}{\cal H}({\bf
k})|\tilde{\Phi}({\bf k})|^2,
\end{equation}
where $\tilde{\Phi}({\bf k})$ is the Fourier transformation of
$\Phi({\bf x})$ and the function ${\cal H}$ is given by
\begin{equation}
{\cal H}({\bf k})=\frac{2}{g}-\sum_\sigma\int{d^3{\bf p}\over
(2\pi)^3}\frac{1-f(\epsilon_+^\sigma)-f(\epsilon_-^\sigma)}{2\epsilon_\mu+|{\bf
k}|^2/4m}
\end{equation}
with $\epsilon_\mu=\epsilon_p-\mu$ and $\epsilon_\pm^\sigma =({\bf
p}\pm{\bf k}/2)^2/(2m)-\mu_\sigma$. Note that ${\cal H}$ is an
even function of ${\bf k}$, it depends only on the amplitude
$k=|{\bf k}|$, ${\cal H}(k)$. For a symmetric system with
$\alpha=0$, the sufficient and complete requirement for stable
normal state is simply expressed as ${\cal H}(0)>0$ and the
critical temperature is determined by ${\cal H}(0)=0$, because of
the condition ${\cal H}(k)\ge {\cal H}(0)$. However, this
requirement becomes incomplete for an asymmetric system with
$\alpha\ne 0$, since $k=0$ is no longer the minimum of the
function ${\cal H}(k)$. While the equality ${\cal H}(0)=0$ gives
two critical temperatures $T_o$ and $T_c$, and the normal phase
below $T_o$ is free from the ordinary BCS instability,
characterized by ${\cal H}(0)>0$, we should demand ${\cal H}(k)>0$
for all $k$ to achieve a real stable normal state, and the
critical temperature should be the maximum of those determined by
${\cal H}(k)=0$. When we calculate the function ${\cal H}(k)$ in
the normal phase, for the region close to the critical temperature
$T_o$, ${\cal H}(k)$ is positive at $k=0$ but becomes negative in
a range of nonzero $ k$. This means that the normal phase there is
stable against homogeneous superfluidity but unstable against
inhomogeneous superfluidity. If we consider only homogeneous
phases, we do find the strange intermediate temperature
superfluidity. However, this is unrealistic when an inhomogeneous
condensed phase enters.

\section {Summary}
\label{s6}
We have investigated the temperature behavior of the homogeneous
and inhomogeneous superfluids in a two-component Fermi gas
with density imbalance. The main conclusions are:\\
(1)For homogeneous superfluid, while the most favored pairing
temperature is nonzero, and in the case with large asymmetry the
superfluidity starts even at finite temperature, the superfluid
density and the number susceptibility are negative at low
temperature $T<T_1$. Therefore, the homogeneous superfluid is
stable only at high temperature $T>T_1$ and unstable at low
temperature $T<T_1$. \\
(2)The LOFF phase is energetically more favored than the
homogeneous superfluid in the region $T<T_1$, and the phase
transition from LOFF to homogeneous superfluid characterized by
the pair momentum happens at $T_1$. Due to the existence of the
inhomogeneous state at low temperature, the so called intermediate
temperature superfluidity\cite{sedrakian,liao,sedrakian3,chen}
disappears and the pairing gap $\Delta$ is a
monotonous function of the temperature.\\
(3)The phase separation is also energetically favored in the
region where the homogeneous superfluid is unstable. From the
calculation of the LOFF number susceptibility, the LOFF phase is
stable only at large asymmetry, while the phase separation should
be the ground state at small asymmetry.\\
(4)When the asymmetry is larger than the maximum value for the
homogeneous superfluid and less than the limit for the LOFF
superfluid, the homogeneous superfluid disappears at any
temperature, but the LOFF superfluid survives and is stable at low
temperature. When the asymmetry is too large, any superfluid
vanishes, and the only possible state is the normal state.\\
(5)The phase transition from homogeneous superfluid or LOFF
superfluid to normal state is of second order, and the transitions
from phase separation to homogeneous superfluid and to LOFF
superfluid are of first order. The four phases meet at the
tetracritical point F. The obtained phase diagram in $T-\alpha$
plane with fixed densities is qualitatively different from the one
in $T-\delta\mu$ plane with fixed chemical potentials.

We only analyzed the stability of homogeneous superfluid and LOFF
superfluid against the phase separation and neglected the
contribution from the surface energy. In further works a detailed
calculation of the phase separation and the consideration of the
surface energy\cite{silva,caldas2} are needed.

Finally we should point out that our result obtained in the simple
but general two-component model for atomic Fermi gas can be
extended and applied to other physical systems, such as isospin
asymmetric nuclear matter with neutron-proton
pairing\cite{sedrakian,Isayev,sedrakian2} and neutral color
superconducting quark matter\cite{shovkovy,alford}.

{\bf Acknowledgement:} The work was supported by the grants
NSFC10428510, 10435080, 10575058 and SRFDP20040003103. We thank
Q.Chen, H.Hu, M.Huang, T.Mizushima, H.Ren, M.Ruggieri and
A.Sedrakian for valuable discussions.

\end{document}